\documentclass[iop,apj]{emulateapj}

\newcommand{\myemail}{agnese.del-moro@durham.ac.uk}
\usepackage{times}
\usepackage{epsfig} 
\usepackage{amssymb}
\usepackage{graphics}
\usepackage{natbib} 

\newcommand{\cgs}{erg~cm$^{-2}$~s$^{-1}$}  
\newcommand{\ergs}{erg~s$^{-1}$} 
\newcommand{\ch}{{\it Chandra}}
\newcommand{\xmmn}{{\it XMM-Newton}}
\newcommand{\xmm}{{\it XMM}}
\newcommand{\spz}{{\it Spitzer}}
\newcommand{\her}{{\it Herschel}} 
\newcommand{\nus}{{\it NuSTAR}} 

\shorttitle{Reflection in a $z\approx2$ heavily obscured quasar}
\shortauthors{A.~Del Moro et al.}

\begin{document}

\title{NuSTAR J033202--2746.8: direct constraints on the Compton reflection in a heavily obscured quasar at $z\approx2$}


\author{A.~Del~Moro\altaffilmark{1}, J.~R.~Mullaney\altaffilmark{2,1}, D.~M.~Alexander\altaffilmark{1}, A.~Comastri\altaffilmark{3}, F.~E.~Bauer\altaffilmark{4,5}, E. Treister\altaffilmark{6}, D. Stern\altaffilmark{7}, F.~Civano\altaffilmark{8,9}, P.~Ranalli\altaffilmark{10}, C.~Vignali\altaffilmark{11,3}, J.~A.~Aird\altaffilmark{1}, D.~R.~Ballantyne\altaffilmark{12}, M.~Balokovi{\' c}\altaffilmark{13}, S.~E.~Boggs\altaffilmark{14}, W.~N.~Brandt\altaffilmark{15,16}, F.~E.~Christensen\altaffilmark{17}, W.~W.~Craig\altaffilmark{18}, P.~Gandhi\altaffilmark{1}, R. Gilli\altaffilmark{3}, C.~J.~Hailey\altaffilmark{19}, F.~A.~Harrison\altaffilmark{13}, R.~C.~Hickox\altaffilmark{8}, S.~M.~LaMassa\altaffilmark{20}, G.~B.~Lansbury\altaffilmark{1}, B.~Luo\altaffilmark{15,16}, S.~Puccetti\altaffilmark{21,22}, M.~Urry\altaffilmark{20} and W.~W.~Zhang\altaffilmark{23}}

\affil{$^1$ Department of Physics, Durham University, South Road, Durham, DH1 3LE, UK; email: \myemail}
\affil{$^2$ Department of Physics \& Astronomy, University of Sheffield, Sheffield, S3 7RH, UK}
\affil{$^3$ INAF - Osservatorio Astronomico di Bologna, Via Ranzani 1, I-40127 Bologna, Italy}
\affil{$^4$ Instituto de Astrof\'{\i}sica, Facultad de F\'{i}sica, Pontificia Universidad Cat\'{o}lica de Chile, 306, Santiago 22, Chile}
\affil{$^5$ Space Science Institute, 4750 Walnut Street, Suite 205, Boulder, Colorado 80301}
\affil{$^6$ Departamento de Astronom\'{i}a, Universidad de Concepci\'{o}n, Casilla 160-C, Concepci\'{o}n, Chile}
\affil{$^7$ Jet Propulsion Laboratory, California Institute of Technology, 4800 Oak Grove Drive, Mail Stop 169-221, Pasadena, CA 91109, USA}
\affil{$^8$ Department of Physics and Astronomy, Dartmouth College, 6127 Wilder Laboratory, Hanover, NH 03755, USA}
\affil{$^9$ Harvard-Smithsonian Center for Astrophysics, 60 Garden Street, Cambridge, MA 02138, USA}
\affil{$^{10}$ National Observatory of Athens, Institute of Astronomy, Astrophysics, Space Applications and Remote Sensing, Metaxa \& Pavlou St., 15236 Penteli, Greece}
\affil{$^{11}$ Universit\`{a} di Bologna, Dipartimento di Fisica e Astronomia, via Berti Pichat 6/2, 40127 Bologna, Italy}
\affil{$^{12}$ Center for Relativistic Astrophysics, School of Physics, Georgia Institute of Technology, Atlanta, GA 30332, USA}
\affil{$^{13}$ Cahill Center for Astrophysics, 1216 East California Boulevard, California Institute of Technology, Pasadena, CA 91125, USA} 
\affil{$^{14}$ Space Sciences Laboratory, University of California, Berkeley, CA 94720, USA}
\affil{$^{15}$ Department of Astronomy and Astrophysics, 525 Davey Lab, The Pennsylvania State University, University Park, PA 16802, USA}
\affil{$^{16}$ Institute for Gravitation and the Cosmos, The Pennsylvania State University, University Park, PA 16802, USA}
\affil{$^{17}$ DTU Space-National Space Institute, Technical University of Denmark, Elektrovej 327, DK-2800 Lyngby, Denmark}
\affil{$^{18}$ Lawrence Livermore National Laboratory, Livermore, CA 94550, USA}
\affil{$^{19}$ Columbia Astrophysics Laboratory, 550 W 120th Street, Columbia University, NY 10027, USA}
\affil{$^{20}$ Yale Center for Astronomy \& Astrophysics, Yale University, Physics Department, PO Box 208120, New Haven, CT 06520-8120, USA}
\affil{$^{21}$ ASI-Science Data Center, via Galileo Galilei, I-00044 Frascati, Italy}
\affil{$^{22}$ INAF-Osservatorio Astronomico di Roma, via Frascati 33, I-00040 Monteporzio Catone, Italy}
\affil{$^{23}$ NASA Goddard Space Flight Center, Greenbelt, MD 20771, USA}

\begin{abstract}
We report \nus\ observations of NuSTAR J033202--2746.8, a heavily obscured, radio-loud quasar detected in the Extended \ch\ Deep Field-South, the deepest layer of the \nus\ extragalactic survey ($\sim$400 ks, at its deepest). NuSTAR J033202--2746.8 is reliably detected by \nus\ only at $E>8$ keV and has a very flat spectral slope in the \nus\ energy band ($\Gamma=0.55^{+0.62}_{-0.64}$; $3-30$ keV). Combining the \nus\ data with extremely deep observations by \ch\ and \xmmn\ (4 Ms and 3 Ms, respectively), we constrain the broad-band X-ray spectrum of NuSTAR J033202--2746.8, indicating that this source is a heavily obscured quasar ($N_{\rm H}=5.6^{+0.9}_{-0.8}\times10^{23}$ cm$^{-2}$) with luminosity $L_{\rm 10-40\ keV}\approx6.4\times10^{44}$ erg~s$^{-1}$. Although existing optical and near-infrared (near-IR) data, as well as follow-up spectroscopy with the Keck and VLT telescopes, failed to provide a secure redshift identification for NuSTAR J033202--2746.8, we reliably constrain the redshift $z=2.00\pm0.04$ from the X-ray spectral features (primarily from the iron K edge). The \nus\ spectrum shows a significant reflection component ($R=0.55^{+0.44}_{-0.37}$), which was not constrained by previous analyses of \ch\ and \xmmn\ data alone. The measured reflection fraction is higher than the $R\sim0$ typically observed in bright radio-loud quasars such as NuSTAR J033202--2746.8, which has $L_{\rm 1.4\ GHz}\approx10^{27}$ W~Hz$^{-1}$. Constraining the spectral shape of AGN, including bright quasars, is very important for understanding the AGN population, and can have a strong impact on the modeling of the X-ray background. Our results show the importance of \nus\ in investigating the broad-band spectral properties of quasars out to high redshift.
\end{abstract}




\section{Introduction}
Many studies in the past 50 years have been devoted to understanding the origin of the X-ray background (XRB) radiation since its discovery in the early 1960's \citep{giacconi1962}. It is now clear that this radiation is due to the emission from individual X-ray sources, with Active Galactic Nuclei (AGN) being the main contributors to the overall XRB emission. The XRB spectrum, as measured from several past and current X-ray missions (e.g., {\it HEAO-1 A2},  \citealt{gruber1992,gruber1999}; {\it BeppoSAX}, \citealt{vecchi1999}; {\it ASCA}, \citealt{gendreau1995,kushino2002}; {\it Swift}-BAT, \citealt{ajello2008}), peaks at $\approx20-30$ keV. Many studies infer that a large population of heavily obscured and Compton-thick AGN (with column densities of $N_{\rm H}>10^{24}$ cm$^{-2}$) are needed to produce this peak (e.g., \citealt{comastri1995,worsley2005,treister2005,ballantyne2006,gilli2007,treister2009}). However, there are still uncertainties regarding the relative contribution of obscured and Compton-thick AGN populations to the XRB spectrum.

Deep X-ray surveys from the \ch\  and \xmmn\  observatories (e.g., \citealt{alexander2003, hasinger2007, comastri2011, xue2011, ranalli2013}) have provided the best constraints on the source population dominating the X-ray emission at $E<10$ keV, allowing us to resolve $\approx70-90$\% of the XRB at energies $E\approx0.5-10$ keV (e.g., \citealt{worsley2005, hickox2006, xue2012}). However, due to their $\approx0.1-10$ keV bandpass, \ch\  and \xmmn\ cannot provide a clear indication of the population dominating at higher energies ($E>10$ keV); this also means that these observatories are relatively insensitive to the identification of the most heavily obscured AGN, where the low-energy emission is suppressed by large column density gas. On the other hand, until recently the X-ray telescopes sensitive at energies $E>10$ keV (e.g., {\it Swift}-BAT, {\it INTEGRAL} and {\it Suzaku}) have yielded direct constraints of only $\approx1-2$\% of the hard X-ray population contributing to the XRB emission at its peak (e.g., \citealt{krivonos2007,ajello2008,bottacini2012}) due to their inherently high background levels. As a consequence, there are still large uncertainties on the predictions from the models of the XRB, which vary significantly depending on the assumed distribution of absorbing column densities, the intrinsic X-ray spectral properties, and the fraction of heavily obscured and Compton-thick AGN (e.g., \citealt{gilli2007,treister2009,ballantyne2011,akylas2012}).

Today, great improvements can be made in measuring the composition of the XRB thanks to the {\it Nuclear Spectroscopic Telescope Array} (\nus; \citealt{harrisonf2013}). \nus\ is the first high-energy orbiting observatory ($E\approx3-79$ keV) equipped with focussing optics, which make this satellite $\approx$2 orders of magnitude more sensitive than the previous-generation hard X-ray ($E>10$ keV) observatories, with one order of magnitude higher angular resolution. With its unique characteristics \nus\ allows us to: 1) identify sources almost independently from their level of obscuration (at least for column densities $N_{\rm H}\lesssim10^{25}$ cm$^{-2}$), therefore overcoming the limitations of the lower-energy observatories currently available; 2) measure the composition of the XRB at its peak energies, providing direct constraints on the contribution from different AGN populations; 3) characterize the broad-band X-ray spectra of AGN, removing ambiguities on the source properties (which are often present when only $E<10$ keV spectra are available), yielding unprecedented constraints on the spectral models (e.g., \citealt{risaliti2013}) even for heavily obscured and Compton-thick AGN out to high redshift ($z\approx2$). 

In this paper we investigate the case of NuSTAR J033202--2746.8, which is detected in the \nus\ observations of Extended \ch\ Deep Field-South (E-CDF-S; Mullaney et al., in prep.). NuSTAR J033202--2746.8 is only significantly detected at $E>8$ keV, showing the hardest band ratio among all of the \nus\ sources detected in the E-CDF-S field so far. This suggests that the source is heavily obscured. Although previously studied at low energies with \ch\ and \xmmn, the X-ray spectrum of this source has never been accurately characterized. Indeed, with the \nus\ data that allow us to constrain its spectral parameters over a broad energy range, we identify a significant reflection component contributing to the spectrum at high energies. Furthermore, using deep infrared (IR) data available in this field, we independently infer the intrinsic power of the AGN, as well as characterize its host galaxy by means of detailed Spectra Energy Distribution (SED) analysis.  

The paper is organized as follows: in section \ref{data} we briefly describe the \nus\ data reduction as well as the lower energy data from \ch\ and \xmmn, and the multi-wavelength data available for the source; in section \ref{src} we report what was known about NuSTAR J033202--2746.8 from the literature; in section \ref{spec} a detailed X-ray spectral analysis is presented, initially using \nus, \ch\  and \xmmn\ data separately, and subsequently in the broad-band $E\approx0.5-30$ keV energy range, jointly fitting the three datasets. In section \ref{ir} we investigate the IR and radio emission of the source through SED decomposition analysis; discussion and conclusions are in sections \ref{discus} and \ref{concl}, respectively.
\begin{figure*}
\centerline{
\includegraphics[scale=0.5]{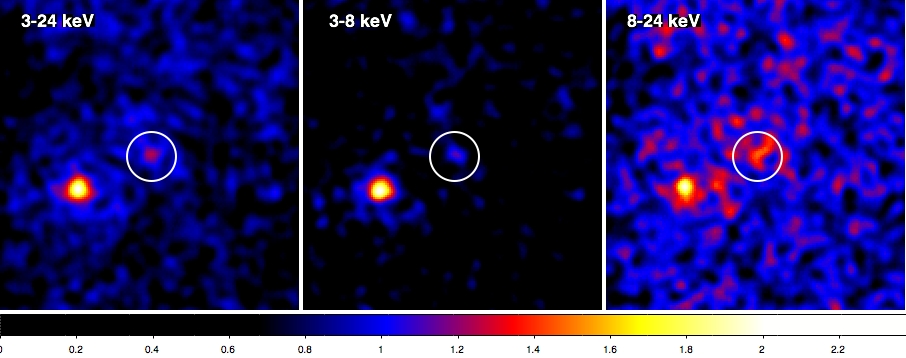}}
\caption{\nus\ smoothed images of the source NuSTAR J033202--2746.8 at $3-24$ keV (left), $3-8$ keV (center), and $8-24$ keV (right), from the FPMA and FPMB modules combined. The white circles have 30$''$ radius. The source is undetected down to a probability $Prob=4\times10^{-4}$ in the $3-8$ keV band, but has a clear detection in the broader $3-24$ keV energy band ($Prob=10^{-9}$), and in the $8-24$ keV band ($Prob\approx4\times10^{-6}$).}
\label{fig.1}
\end{figure*}
Throughout the paper we assume a cosmological model with $H_0=70\ \rm km\ s^{-1}\ Mpc^{-1}$, $\Omega_M=0.27$ and $\Omega_{\Lambda}=0.73$ \citep{spergel2003}. All the errors are quoted at a 90\% confidence level, unless otherwise specified. 

\section{Data}\label{data}
\subsection{\nus\ data}\label{ndata}
The \nus\ satellite is equipped with two telescopes, which focus X-ray photons onto two independent $\approx$12$\times$12 arcmin$^2$ focal planes, called Focal Plane Modules (FPMA and FPMB). \nus\ is sensitive to hard X-rays in the energy range $E\approx3-79$ keV and has an unprecedented angular resolution at these energies of 18$''$ FWHM with a half power diameter of 58$''$, and an energy resolution of 400 eV FWHM at 10 keV and 0.9 keV at 60 keV (\citealt{harrisonf2013}).

The E-CDF-S, where NuSTAR J033202--2746.8 is detected, has been observed by \nus\ as part of the extragalactic survey program. The extragalactic survey is designed as three components (see Table 6 of \citealt{harrisonf2013}): a deep small-area survey in the E-CDF-S field (now also including the Extended Groth Strip, EGS), a medium wider-area survey in the Cosmic Evolution Survey (COSMOS; \citealt{scoville2007}) field, and a large area serendipitous survey conducted in the fields of other targeted \nus\ observations \citep{alexander2013}, including $\approx$100 {\it Swift}-BAT identified AGN. 
The deep survey of the E-CDF-S field is currently composed of two passes of sixteen 50-ks observations each, covering $\approx$0.3 deg$^2$. The observations were completed in April 2013: the first pass observations were performed between October and November 2012; the second pass observations were performed between March and April 2013 (Mullaney et al., in prep.). 

The \nus\ data were processed using the \nus\ Data Analysis Software (NuSTARDAS) package (v.1.2.0) and the NASA's HEASARC software (HEAsoft v.6.14\footnote{\url{http://heasarc.nasa.gov/lheasoft/}}). The cleaned event files were produced through standard pipeline filtering using the \texttt{nupipeline} task, and the latest calibration files available in the \nus\ Calibration Database (CALDB; v.20130315). Science images, background maps and vignetting-corrected exposure maps were then produced in three energy bands: $3-8$ keV, $8-24$ keV and $3-24$ keV (see also \citealt{alexander2013}), using standard NuSTARDAS tools and customized scripts (\texttt{nuskybgd}; Wik et al., in prep.), and mosaiced using the XIMAGE graphical tool. 

\subsubsection{Source detection}\label{det}
To detect sources in the \nus\ $3-24$ keV mosaic images and obtain their \nus\ positions, we initially used the CIAO tool \texttt{wavdetect}, including exposure maps and background maps (see Sect. \ref{ndata}). We used a low probability threshold of $10^{-4}$ and wavelet scales $=[4, 5.66, 8, 11.31, 16]$ pixels to ensure a detection in at least one of the two \nus\ FPMs.\footnote{We note that the detection method applied here is different from the detection technique adopted in the \nus\ E-CDF-S catalogue paper (Mullaney et al., in prep.). However, the position and detection reliability obtained for NuSTAR J033202--2746.8 are consistent between the two methods.} 
NuSTAR J033202--2746.8 is detected by \texttt{wavdetect} in the FPMA image only, with coordinates: RA$=53.0083$ deg and Dec=$-27.7807$ deg, and a positional uncertainty of $0.7''$; we note, however, that given the size of the \nus\ PSF, the positional uncertainty given by \texttt{wavdetect} is underestimated. The typical astrometry uncertainties in the \nus\ images are estimated to be $\approx8''$ \citep{harrisonf2013}. 

After several tests of our source detection technique, we ascertained that \texttt{wavdetect} struggles to account for the high background levels that characterize the \nus\ E-CDF-S data. We therefore resorted to aperture photometry, calculating the Poisson false probability to estimate whether the measured counts above the background within a given extraction region constitute a significant detection. We performed the aperture photometry on the mosaic images (for the FPMA and FPMB separately, and then combined) in three energy bands ($3-8$, $8-24$ and $3-24$ keV; see Fig. \ref{fig.1}) using a 30$''$ radius extraction region to determine the total count rates of the source in each band. Such a large extraction region is chosen to account for the \nus\ PSF, and the astrometric uncertainties in each individual observation. To estimate the background counts we performed aperture photometry on the background maps, using the same extraction region as for the source, and calculated the Poisson false probability using the incomplete $\Gamma$ function. We consider the source significantly detected when the Poisson false probability $Prob\le10^{-4}$. The results of this reliability detection approach for NuSTAR J033202--2746.8 are summarized in Table \ref{tab.det}. 

Considering the three \nus\ bands separately, NuSTAR J033202--2746.8 is formally undetected at $E=3-8$ keV in both FPMA and FPMB images, as well as in the combined image, while it is reliably detected in the full \nus\ band in all the mosaics ($3-24$ keV; $Prob=2\times10^{-6}$ in FPMA; $Prob=10^{-5}$ in FPMB, and $Prob=10^{-9}$ in the combined FPMA and FPMB image), and in the hard band in the FPMA, and in the combined FPMA and FPMB mosaic ($8-24$ keV; $Prob\approx5\times10^{-5}$ and $4\times10^{-6}$, respectively; however it is not significantly detected in the FPMB image: $Prob=4\times10^{-3}$). NuSTAR J033202--2746.8 has the hardest band ratios of any of the sources detected by \nus\ in the E-CDF-S field (Mullaney et al., in prep.): $\rm CR(8-24)/CR(3-8)=1.61\pm$0.62 (from the combined FPMA and FPMB data; where CR are the aperture-corrected net count rates in the specified energy bands\footnote{Since the shape of the PSF of \nus\ changes with off-axis angle, a mean aperture correction was calculated for NuSTAR J033202--2746.8 from the modeled PSFs and then applied to the source count rates ($corr=2.20$ for a 30$''$ radius extraction region, and $corr=1.59$ for a 45$''$ radius extraction region; see Sect. \ref{spec}).}; see Table \ref{tab.det}). 

We cross-matched the \nus\ source position with the E-CDF-S \ch\ source catalogue (\citealt{lehmer2005,xue2011}) and the \xmm-CDFS catalogue (\citealt{ranalli2013}) to identify a low-energy X-ray counterpart to the source (see Sect. \ref{lowx}). Due to the positional uncertainties of the \nus\ data we used a search radius of 15$''$. We found one match (CXO~J033201.4--274647; XID 83 in the \citealt{xue2011} catalogue) in the \ch\ catalogue within our searching radius, at $\sim12''$ offset from the \nus\ position; the next nearest \ch\ neighbor lies $>30''$ from the \nus\ centroid. In the \citet{ranalli2013} catalogue we found an \xmm\ counterpart to NuSTAR J033202--2746.8 $\sim13''$ from the \nus\ position (XMMCDFS~J033201.3--274647; XID 214); no other \xmm\ sources lie in the field within 35$''$. We note that there are no other \nus\ sources nearby (within $>$1 arcmin, see below) that could be associated with the identified \ch\ and \xmmn\ counterparts.
\begin{table*}
\begin{center}
\caption{NuSTAR J033202--2746.8 detection summary. \label{tab.det}}
\begin{tabular}{c c c c c c c c}
\hline\hline
\rule[-1.mm]{0pt}{3ex}ID & Module  & Count rate & $Prob$ & Count rate & $Prob$ & Count rate & $Prob$ \\
 (1) & (2) & (3) & (4) & (5) & (6) & (7) & (8)\\
 &     & \multicolumn{2}{c}{\rule{3cm}{0.4pt}} & \multicolumn{2}{c}{\rule{3cm}{0.4pt}} & \multicolumn{2}{c}{\rule{3cm}{0.4pt}} \\
\rule[-1.8mm]{0pt}{3ex}  &     & \multicolumn{2}{c}{$3-24$ keV} &  \multicolumn{2}{c}{$3-8$ keV} & \multicolumn{2}{c}{$8-24$ keV} \\
\tableline    
\rule[-1.5mm]{0pt}{4ex}NuSTAR J033202--2746.8 & FPMA  & 1.074$\pm$0.251 & 2.4$\times10^{-6}$ & 0.329$\pm$0.147 & 9.5$\times10^{-3}$ & 0.692$\pm$0.191 & 4.8$\times10^{-5}$\\
\rule[-1.5mm]{0pt}{3ex}    & FPMB         & 0.927$\pm$0.232 & 1.0$\times10^{-5}$ & 0.354$\pm$0.139 & 3.6$\times10^{-3}$ & 0.428$\pm$0.170 & 4.1$\times10^{-3}$\\
\rule[-1.5mm]{0pt}{3ex}    & FPMA \& FPMB & 0.956$\pm$0.170	& 1.0$\times10^{-9}$ & 0.332$\pm$0.101 & 3.0$\times10^{-4}$ & 0.535$\pm$0.127 & 3.7$\times10^{-6}$\\
\tableline 
\end{tabular}
\end{center}
NOTES: (1) \nus\ source name; (2) \nus\ module; (3), (5) and (7) aperture-corrected net count rates in the $3-24$, $3-8$ and $8-24$ keV bands, respectively, in units of ks$^{-1}$; (4), (6) and (8) Poisson false probability of detection in the $3-24$, $3-8$ and $8-24$ keV bands, respectively. The source is considered detected if $Prob<10^{-4}$.
\end{table*}

\subsubsection{Spectral extraction}
Given the \nus\ mapping strategy on the E-CDF-S field, there are several overlapping regions between various pointings. NuSTAR J033202--2746.8 lies in the field of view of 9 different observations (although it is not detected in individual observations).
The \nus\ spectra were extracted from each individual observation using a circular region of 45$''$ radius, while the background spectra were extracted from four circular source-free regions\footnote{The background regions were selected also avoiding all known \ch\ sources with fluxes brighter than $f_{\rm 2-8\ keV}=5\times10^{-15}$ \cgs, even if they are not detected by \nus.} of 80$''$ each. We note that there are no other \nus\ detected sources within the source extraction region, and the closest \nus\ source lies $\approx1.7$ arcmin away, so we do not expect contamination from nearby sources to be an issue for our source spectra. The spectra were then combined (FPMA and FPMB modules separately) using the HEAsoft tool \texttt{addascaspec} to increase the counting statistics.

\subsection{X-ray data at E$<10$ keV}\label{lowx}

NuSTAR J033202--2746.8 is located in the region of the E-CDF-S covered by the deepest \ch\ (4 Ms; \citealt{xue2011}) and \xmmn\ (\xmm-CDFS, 3 Ms; \citealt{ranalli2013}) data, which provide excellent quality lower energy ($E\approx0.5-10$ keV) spectral information for the source.  
The 4 Ms \ch\ data include 23 observations performed between 1999-2007 (2 Ms; \citealt{luo2008}) and 31 observations performed between March and July 2010 (see Table 1 of \citealt{xue2011}). Details on the observations and data reduction are described in \citet{luo2008} and \citet{xue2011}. Briefly, the data were processed using the \ch\ Interactive Analysis of Observations\footnote{\url{http://cxc.cfa.harvard.edu/ciao/index.html}} (CIAO; version 4.3 and CALDB 4.4.1.; \citealt{fruscione2006}) tools and the {\em ACIS Extract} (AE) software package\footnote{The {\em ACIS Extract} software package and Users Guide are available at \url{http://www.astro.psu.edu/xray/acis/acis\_analysis.html}.} \citep{broos2010,broos2012}. 
The \ch\ spectra of NuSTAR J033202--2746.8 were produced using the AE software (version 2011-03-16), which extracted the spectra in each individual observation from $\sim4.5-5''$ radius regions (enclosing 90\% of the encircled energy), together with the background spectra and relative response matrices and ancillary files and combines them appropriately, calling the FTOOLs \texttt{addrmf} and \texttt{addarf} and using the observation's exposure to calculate the weights. The resulting spectra are corrected for the energy-dependent PSF shape and extraction aperture.

For the \xmmn\ data we used the observations taken between 2001-2002 (PI: J. Bergeron), with a total exposure of 541 ks, and the more recent ultra-deep observations taken between 2008-2010 (PI: A. Comastri), giving a total of 33 observations in each of the three EPIC cameras (PN, MOS1 and MOS2). The data were processed using the standard \xmmn\  Science Analysis Software\footnote{\url{http://xmm.esac.esa.int/sas/}} (SAS; v10.0.0), and filtered for high background flares (see \citealt{ranalli2013}, for details on the reduction). From each observation the spectra of NuSTAR J033202--2746.8 have been extracted from a circular region of 13.75$''$ radius, while the background spectra have been extracted using a 49.5$''$ radius, source-free region (in PN; 40$''$ in MOS1, and 29.5$''$ in MOS2). Corresponding ancillary and response files were also obtained from each observation and each EPIC camera separately. All of the PN, MOS1 and MOS2 spectra (and related files) from the different observations were then combined using \texttt{mathpha}, \texttt{addrmf} and \texttt{addarf} tools, using appropriate weights calculated from the effective exposure in each spectrum. 
Since the MOS1 and MOS2 cameras have similar characteristics, and therefore similar responses, we also summed together the source (and background) spectra extracted from these two cameras, in order to improve the signal-to-noise ratio ($S/N$). 

\subsection{Multi-wavelength data}\label{multiw}

The \ch\ Deep Field-South (CDF-S), in the central region of the E-CDF-S, is one of the most intensely observed fields in the sky, with deep observations available in the optical, IR and radio bands, among others. In particular the central region of CDF-S ($\approx$160 arcmin$^2$) has been observed in the mid-IR with \spz-IRAC (at 3.6, 4.5, 5.8 and 8 $\mu$m; Dickinson et al., in prep.), and MIPS (24 $\mu$m; GOODS-\spz\ Legacy survey; PI: Mark Dickinson), 
 and in the far-IR with \her\ at 100 and 160 $\mu$m as part of the GOODS-\her\ program (GOODS-H; \citealt{elbaz2011}), and 250, 350 and 500 $\mu$m as part of the \her\ {\it Multi-tiered Extragalactic Survey} (HerMES; \citealt{oliver2012}). A wider area of E-CDF-S has been observed at 100 and 160 $\mu$m by the {\it PACS Evolutionary Probe} survey (PEP; \citealt{lutz2011}). In this work we only use the data up to 250 $\mu$m. The depth of these data in the mid- and far-IR reaches $S_{\rm 24}\approx$20 $\mu$Jy, $S_{\rm 100}\approx$0.6 mJy, $S_{\rm 160}\approx$1.3 mJy (3$\sigma$) and $S_{\rm 250}\approx$4.6 mJy (5$\sigma$), where at 100 and 160 $\mu$m the sensitivities are those of the GOODS-H and PEP surveys combined (see \citealt{magnelli2013}), while for the 250 $\mu$m band we quote the HerMES data sensitivity from \citet{oliver2012}. 
 
In the radio band the E-CDF-S field ($\approx$0.3 deg$^2$) has been observed by the Very Large Array (VLA) at 1.4 GHz, with a typical sensitivity of 7.4 $\mu$Jy (5$\sigma$) per 2.8$''\times 1.6''$ beam \citep{miller2013}. For details on the observations and the source catalogues produced in each individual survey in the mid-IR, far-IR and radio bands we refer to \citet{magnelli2011}, \citet{elbaz2011}, \citet{magnelli2013} and \citet{miller2013}. 

\section{NuSTAR J033202--2746.8}\label{src}

\subsection{Previous results}\label{lit}

Since NuSTAR J033202--2746.8 lies in one of the most observed patches of the sky, this source has been included in several previous population studies, especially focussed on the X-ray band at $E<10$ keV. 
Already detected in the \ch\ 1 Ms survey (XID 70; \citealt{giacconi2002}), this source was found to have X-ray fluxes $f_{\rm 0.5-2}\approx6.5\times10^{-16}$ \cgs, and $f_{\rm 2-10}\approx1.2\times10^{-14}$ \cgs, with a hard hardness ratio\footnote{$HR=(H-S)/(H+S)$, where $H$ and $S$ are the count rates in the hard ($2-8$ keV) and soft ($0.5-2$ keV) energy bands, respectively.} $HR=0.47\pm0.04$, and a relatively faint magnitude of the candidate optical counterpart $R=23.62\pm0.12$ mags (\citealt{wolf2004}). These characteristics already suggested obscuration in both the X-ray and optical bands. Indeed, \citet{civano2005} included NuSTAR J033202--2746.8 in their sample of optically faint sources, i.e. sources with high X-ray to optical flux ratios ($\rm X/O$; see also \citealt{delmoro2009}). A more recent work by \citet{mainieri2008} reported an even fainter $R-$band magnitude of $R>25.5$ mag (AB) from the Wide Field Imager (WFI) catalogue from the ESO Imaging Survey (EIS). The X-ray spectral properties of this source based on the \ch\ 1 Ms data, reported by \citet{tozzi2006}, revealed a very hard spectral slope ($\Gamma=0.55^{+0.20}_{-0.20}$) and a modest column density of $N_{\rm H}\approx4\times10^{22}$ cm$^{-2}$ for an assumed photometric redshift of $z=1.07$ (from the COMBO-17 survey; \citealt{wolf2001,wolf2004}). More recent works (e.g., \citealt{xue2011,castellomor2013}) reported a revised photometric redshift for NuSTAR J033202--2746.8, $z=1.499$, based on the MUSYC survey (\citealt{luo2010}). In the studies by \citet{castellomor2013} and \citet{georgantopoulos2013}, which are based on the ultra-deep \xmmn\ data (\citealt{ranalli2013}), NuSTAR J033202--2746.8 is identified as a heavily obscured AGN and has a higher column density ($N_{\rm H}\approx(2-3)\times10^{23}$ cm$^{-2}$) than what was previously measured, but still with a hard intrinsic power-law slope\footnote{Note that these two analyses use different XSPEC spectral models to constrain the spectral parameters of their sources.} of $\Gamma\approx1.2$ in \citet{castellomor2013} and $\Gamma\approx0.9$ in \citet{georgantopoulos2013}. However, these two works used different redshifts for the source: $z=1.499$ in \citet{castellomor2013} and $z=2.0$ in \citet{georgantopoulos2013}, estimated from the \xmm\ X-ray spectrum. 

NuSTAR J033202--2746.8 has also been detected as a bright radio source, with $S_{\rm 1.4\ GHz}=53.6\pm0.05$ mJy (e.g., \citealt{bonzini2012,miller2013}). \citet{miller2013} identified a complex radio morphology for this source (VLA~J033201.4$-$274648), consisting of two overlapping lobes (see also \citealt{kellerman2008}). 

In summary, although NuSTAR J033202--2746.8 is included in several spectroscopic and photometric redshift surveys (e.g., \citealt{wolf2001,wolf2004,szokoly2004,zheng2004,cardamone2010}) a unique and robust redshift identification has not been found in these studies and has yet to be unambiguously determined. Moreover, despite several previous studies using the deepest available X-ray data from \ch\  and \xmmn\ having included NuSTAR J033202--2746.8 in their samples, an accurate characterization of the source spectrum is still lacking. Indeed, the spectral results obtained for this source, in particular the very flat photon index compared to those typically found for unobscured AGN ($\Gamma=1.8\pm0.2$; e.g., \citealt{nandra1994,mainieri2002,caccianiga2004,mateos2005b,tozzi2006,burlon2011}) suggests that the X-ray spectrum of NuSTAR J033202--2746.8 is likely to be more complex than what has been considered so far. 
 
\subsection{Keck and VLT-XSHOOTER follow-up spectra}\label{opt}

Due to the uncertainties on the source redshift from existing optical and IR data, we performed follow-up spectroscopic observations in the optical band with Keck and in the broad-band ultraviolet-to-infrared wavelengths with the VLT XSHOOTER spectrograph (\citealt{vernet2011}). 
The Keck observations were performed on the nights of the 4th and 5th of October 2013. NuSTAR J033202--2746.8 was observed with the dual-beam Low Resolution Imaging Spectrometer (LRIS; \citealt{oke1995}) for a total exposure of 3.5~h. The first night had variable conditions, while the second night was photometric. We observed through a slitmask with $1.5''$ wide slitlets, using the 400/3400 $\ell/$mm grism on the blue arm, the 400/8500 $\ell/$mm grating on the red arm, and the 5600 \AA\ dichroic to split the light. This instrument configuration provides sensitivity across the complete optical window, from $\sim$3200 \AA\ to $\sim$1 $\mu$m. The data reduction was performed using standard procedures and rely on the best 2.5~h of integration.  

The VLT observations were performed on the 3rd of November 2013 (PI.: E. Treister) using the UVB ($300-559.5$ nm), VIS ($559.5-1024$ nm) and NIR ($1024-2480$ nm) spectroscopic arms of XSHOOTER to cover the broadest possible wavelength range. A long slit of $0.9''$ width was used in the VIS and NIR arms. In order to reduce the background level, the observations were split into $12\times300$ s for the NIR data, for a total on-source exposure of 1 hour, while in the UVB and VIS arms the observations were split into $12\times163$ s (0.54 h in total) and $12\times230$ s (0.77 h in total), respectively. The sky conditions were clear, with $0.5''$ seeing. The data were reduced using the standard ESO XSHOOTER pipelines (v.2.3.0) and calibrations. 

NuSTAR J033202--2746.8 was not detected in the Keck optical spectroscopic observations, nor in the UV and optical bands with VLT, with no sign of either emission lines nor of continuum in these data. This is not unexpected since the source has already been reported to be very faint in the optical (Sect. \ref{lit}), and thus probably very reddened in these bands. From the Keck-LRIS data we estimate a line flux upper limit of a few $\times 10^{-18}$ \cgs. In the near-IR band the VLT-XSHOOTER spectrum shows a faint continuum emission, detected at $S/N>3$ only over limited intervals. Moreover, no emission lines are detected in the near-IR spectrum. Assuming our best estimate of the source redshift from the X-ray spectra, $z\approx2.0$ (see Sect. \ref{spec}), we place upper limits on the flux of the emission lines we expect to see in the near-IR band, such as [OIII]$\ \lambda 5007$ \AA\ (15012 \AA, observed frame) and H$\alpha\ \lambda6563$ \AA\ (19689 \AA, observed frame). Using a Gaussian line profile with full width half maximum $\rm FWHM=500$ km~s$^{-1}$, we integrated the spectrum over the wavelength range expected for the [OIII] and H$\alpha$ emission lines and we obtain: $f_{\rm [OIII]}<8.2\times10^{-18}$ \cgs\ and $f_{\rm H\alpha}<4.8\times10^{-18}$ \cgs\ (3$\sigma$ upper limits). We note, however, that the exact sensitivity limit is wavelength dependent, depending on the telluric emission lines, the line width, atmospheric transmission, and instrumental parameters such as the choice of dichroic, as well as the CCD sensitivities. It is also important to mention that the detected near-IR continuum is fainter than expected from the $K-$band magnitude of NuSTAR J033202--2746.8 ($K\approx20.1$ mag, AB; e.g., \citealt{luo2010,xue2011}). It is therefore possible that significant flux losses may have affected the spectrum, possibly due to uncertainty on the pointing or slit position during the acquisition. Although it is not possible to verify whether this is the case, nor quantify the extent of the losses, we advise that the line flux upper limits reported above might be underestimated. 

\section{Spectral analysis and results}\label{spec}
\subsection{The \nus\ spectra}\label{nuspec}
The \nus\ spectra for NuSTAR J033202--2746.8 have an effective exposure time $t\sim3.7\times10^{5}$ sec (in each FPM), and net, aperture-corrected count rates $CR(3-30)=0.86\pm 0.16$ ks$^{-1}$ (FPMA) and $CR(3-30)=0.37\pm 0.16$ ks$^{-1}$ (FPMB). We note that the source is detected in FPMB with lower reliability than in FPMA (formally undetected according to our false probability threshold, see Sect. \ref{det}), probably due to higher background fluctuations, which might explain the discrepant net count rates measured from the spectra in the two FPMs. The total source counts in the spectra are largely dominated by the background, which accounts for $\approx$86-93\% of the total count rate at $E=3-30$ keV, with only $\approx200$ net counts (FPMA; $\approx100$ in FPMB) coming from the source. We only consider here data up to $E=30$ keV (observed frame) because above these energies the instrumental background dominates by a large factor over the source spectrum (e.g., \citealt{harrisonf2013}), yielding large uncertainties on the signal-to-noise ratio and the source spectral shape. Due to the limited source counting statistics, we perform the spectral fitting of the source and background spectra together using Cash statistics (C-stat; \citealt{cash1979}). Since the background is the main contributing component in our spectra, we need to carefully model its spectrum before attempting to fit the source $+$ background. 
Therefore, we initially fit the background spectrum only, binning the data with a minimum of 50 cts/bin and using $\chi^2$ statistics to find the best-fit model parameters; we fit the background spectrum using the model adopted by Wik et al. (in prep.). We then analyzed the \nus\ source ($+$ background) spectra using C-stat, assuming a simple power-law model for the source (including Galactic absorption $N_{\rm H,\ Gal}=9.0\times10^{19}$ cm$^{-2}$; \citealt{dickey1990}), plus the background model with all the parameters fixed to their best-fit values previously obtained, and rescaled proportionally to the source/background extraction areas. 
The photon index resulting from this initial spectral fit is $\Gamma=0.55^{+0.62}_{-0.64}$, which is flat compared to the typical intrinsic spectral slope of AGN observed at these energies ($\Gamma\approx1.8-2.0$; e.g., \citealt{burlon2011,alexander2013}). This indicates that the spectrum of NuSTAR J033202--2746.8 is rising to high energies, as also hinted by the extreme band ratio (Sect. \ref{det}), suggesting the presence of strong absorption and possibly reflection of the intrinsic nuclear emission. 
\begin{table}
\begin{center}
\caption{X-ray spectral models.\label{tab.mo}}
\begin{tabular}{c c}
\hline
\hline
\rule[-1.7mm]{0pt}{4ex}Models & XSPEC components\\
\tableline
\rule[-1.5mm]{0pt}{4ex}Model 1 & WABS $\times$ (PO $+$ ZWABS $\times$ (PO $+$ ZGAUSS)) \\
\rule[-1.5mm]{0pt}{3ex}Model 2 & WABS $\times$ (PO $+$ ZWABS $\times$ (PO $+$ ZGAUSS) $+$ PEXRAV)\\
\rule[-1.5mm]{0pt}{3ex}Model 3 & WABS $\times$ (PO $+$ ZWABS $\times$ (ZGAUSS $+$ PEXRAV))\\
\tableline
\end{tabular}
\end{center}
\end{table}%

\subsection{Spectral constraints from \ch\ and \xmmn}\label{lowxspec}

\begin{table*}
\begin{center}
\caption{Best-fit spectral parameters of the join fit \nus, \xmmn\ and \ch\  spectra for the different models.\label{tab.sp}}
\begin{tabular}{l c c c}
\hline
\hline
\rule[-1.5mm]{0pt}{4ex}Parameters & Model 1 & Model 2 & Model 3  \\
\tableline
\rule[-1.5mm]{0pt}{4ex}$\Gamma$              & 1.47$^{+0.18}_{-0.18}$ (1.8 f$^a$)             & 1.59$^{+0.21}_{-0.21}$ (1.8 f)                 & 2.68$^{+0.12}_{-0.13}$ (1.8 f)\\
\rule[-1.5mm]{0pt}{3ex}$N_{\rm H}^b$         & 52.3$^{+8.1}_{-7.9}$   (65.0$^{+4.5}_{-4.4}$)  & 55.6$^{+9.4}_{-8.2}$   (64.0$^{+4.9}_{-4.7}$)  & --  \\
\rule[-1.5mm]{0pt}{3ex}$N_{\rm H}^c$         & --                                             &  --                              & 37.5$^{+5.5}_{-5.6}$ (10.9$^{+2.2}_{-2.3}$)\\
\rule[-1.5mm]{0pt}{3ex}R                     & --                                             & $-$1.0 f                                       & $-$1.0 f               \\
\rule[-1.5mm]{0pt}{3ex}$z$                   & 2.00$^{+0.05}_{-0.06}$ (2.01$^{+0.05}_{-0.05}$)& 2.00$^{+0.04}_{-0.04}$ (2.00$^{+0.04}_{-0.04}$)& 2.01$^{+0.04}_{-0.04}$(2.00$^{+0.03}_{-0.03}$) \\
\rule[-1.5mm]{0pt}{3ex}EW$^d$ (Fe K$\alpha$) & 123$^{+102}_{-78}$     ($<155$)                & 144$^{+90}_{-90}$      (79$^{+90}_{-69}$)      & $<65$ (236$^{+128}_{-100}$)\\
\rule[-1.5mm]{0pt}{3ex}$f_{scatt}^e$         & 5.3\%                  (3.1\%)                 & 4.4\%                  (2.8\%)                 &  --                   \\
\rule[-1.5mm]{0pt}{3ex}$C_1^f$ (FPMA)        & 1.0 f                                          & 1.0 f                                          &  1.0 f                \\
\rule[-1.5mm]{0pt}{3ex}$C_2^f$ (FPMB)          & 0.96$^{+0.06}_{-0.06}$ (0.97$^{+0.06}_{-0.06}$)& 0.97$^{+0.06}_{-0.06}$ (0.97$^{+0.06}_{-0.06}$)& 0.99$^{+0.06}_{-0.06}$ (0.98$^{+0.06}_{-0.06}$)  \\
\rule[-1.5mm]{0pt}{3ex}$C_3^f$ (\xmm-PN)        & 1.07$^{+0.48}_{-0.27}$ (1.00$^{+0.48}_{-0.26}$)& 1.08$^{+0.50}_{-0.27}$ (1.07$^{+0.51}_{-0.27}$)& 1.18$^{+0.64}_{-0.32}$ (1.52$^{+0.70}_{-0.38}$)  \\
\rule[-1.5mm]{0pt}{3ex}$C_4^f$ (\xmm-MOS)       & 0.59$^{+0.27}_{-0.15}$ (0.55$^{+0.27}_{-0.14}$)& 0.60$^{+0.27}_{-0.15}$ (0.59$^{+0.28}_{-0.15}$)& 0.66$^{+0.31}_{-0.18}$ (0.88$^{+0.41}_{-0.21}$)  \\
\rule[-1.5mm]{0pt}{3ex}$C_5^f$ (\ch)       & 1.07$^{+0.48}_{-0.27}$ (1.00$^{+0.48}_{-0.25}$)& 1.08$^{+0.49}_{-0.27}$ (1.07$^{+0.50}_{-0.27}$)& 1.20$^{+0.66}_{-0.33}$ (1.59$^{+0.71}_{-0.38}$)  \\
\rule[-1.5mm]{0pt}{3ex}C-stat                & 3003.5/3373            (3012.5/3374)           & 3002.7/3372            (3004.5/3373)          & 3036.6/3373 (3177.1/3374)  \\
\tableline
\end{tabular}
\end{center}
NOTES:$^a$ An ``f'' next to a parameter means that the parameter was fixed during the fit; $^b$ hydrogen column density of the transmitted component in units of $10^{22}$ cm$^{-2}$; $^c$ hydrogen column density of the reflected component in units of $10^{22}$ cm$^{-2}$; $^d$ rest-frame equivalent width of the iron emission line (1$\sigma$ width fixed at 50 eV) in units of eV; $^e$ fraction of the intrinsic emission scattered in the soft band ($E<2$ keV); $^f$ cross-calibration factors between the \nus\ FPMs and the other observatories and instruments. 
\end{table*}%

Since NuSTAR J033202--2746.8 is formally undetected at $E<8$ keV by \nus, the spectral information provided by the \nus\ data is not sufficient to fully characterize the properties of NuSTAR J033202--2746.8. In particular, we need the lower energy data from \ch\ and \xmmn\ to constrain the photoelectric absorption cutoff energy as a probe of the amount of obscuration of the nuclear emission. Since we cannot obtain a redshift measurement from our spectroscopic follow-up observations, we also attempt to measure the source redshift through the X-ray spectral features (e.g., \citealt{iwasawa2012,georgantopoulos2013}).

The \ch\ and \xmm\ spectra have high counting statistics: 2136 cts in \ch, with an effective exposure $t\approx$4.1 Ms, 3938 cts in \xmm-PN ($t\approx$2.8 Ms) and 4019 cts in \xmm-MOS ($t\approx$2.1 Ms). The high number of counts allow us to use $\chi^2$ statistics for the spectral fitting; however, while the background in the \ch\ spectra is negligible ($\approx$4\% of the total count rate), in the \xmm\ spectra the background accounts for more than half of the total spectral counts (52-55\%). We binned the \ch\ spectrum using the AE software (see Sect. \ref{lowx}) to have at least a signal-to-noise ratio $S/N>3$ in each bin. To rebin the \xmm\ spectra we used the SAS tool \texttt{specgroup}, which allows grouping of the net spectral counts (i.e., background subtracted) to have a $S/N>3$ in each bin.

We then fit the \ch\ and \xmm\ data separately to allow for a comparison of the resulting best-fitting parameters.
We use two models in XSPEC to fit these low-energy spectra: Model 1 is composed of a simple power-law model modified by Galactic and intrinsic absorption, and also includes a soft scattered power-law component\footnote{We also tried modeling the soft emission with a collisionally ionized diffuse gas model (APEC in XSPEC; \citealt{smith2001}), which could be gas in the narrow line regions (e.g., \citealt{bianchi2006}). However, the temperature of the gas $kT$, which is the main parameter of the model, could not be constrained. We therefore adopt a simple power law, with $\Gamma$ linked to that of the intrinsic power law.}, and a Gaussian line at rest-frame 6.4 keV; Model 2 includes, in addition, a reflection component (PEXRAV in XSPEC; \citealt{magdziarz1995}), with the spectral slope linked to that of the primary intrinsic power law; we fixed the reflection parameter to $\rm R=-1$ (corresponding to a covering factor of the cold matter to the X-ray source of $\Omega=2\pi$) and the cutoff energy $E_{c}=250$ keV (e.g., \citealt{akylas2012,ballantyne2013}), assuming solar abundances for all elements and an inclination angle of $60^{\circ}$ (e.g., \citealt{ueda2007,corral2011}). The XSPEC components adopted in our models are listed in Table \ref{tab.mo}. We note that these models are widely used to fit both radio-loud and radio-quiet AGN spectra (e.g., \citealt{eracleous2000,reeves2000,hardcastle2006}); we will discuss further implications in Sect. \ref{discus}. We initially fixed the redshift in our models to $z=1.499$ (e.g., \citealt{luo2008}; see Sect. \ref{lit}). However, clear residuals in the spectra between $E\approx2-3$ keV, likely related to the iron K$\alpha$ emission line (rest-frame $E\approx6.4$ keV) and iron absorption edge (rest-frame $E\approx7.1$ keV), suggest that this redshift is not correct. We therefore keep the redshift as a free parameter in our models. 

The \ch\ spectrum of NuSTAR J033202--2746.8 is fitted between 0.5 and 8 keV, and for Model 1 we obtain a flat spectral slope $\Gamma=1.05^{+0.45}_{-0.28}$ (all of the errors on the parameters are estimated at a 90\% confidence level), with intrinsic column density of $N_{\rm H}=(3.5^{+2.0}_{-1.0})\times10^{23}$ cm$^{-2}$ ($\chi^2/d.o.f.=97.9/101$). The redshift for this model is $z=1.89^{+0.16}_{-0.05}$, mainly constrained through the iron K$\alpha$ emission line, with equivalent width EW=$281^{+167}_{-276}$ eV (rest frame), and the iron edge (see Fig. \ref{fig.cont}); this redshift is consistent with the estimate from \citet{georgantopoulos2013}. If we include a reflection component (i.e., Model 2) we obtain a steeper slope $\Gamma=1.65^{+0.22}_{-0.18}$, with $N_{\rm H}=(5.5^{+0.7}_{-0.7})\times10^{23}$ cm$^{-2}$ and $z=1.99^{+0.07}_{-0.08}$ ($\chi^2/d.o.f.=96.0/100$). We then used the Bayesian information criterion ($BIC$; \citealt{schwarz1978}) to verify whether the reflection component significantly improves the fit. However, the resulting $\Delta BIC_{21}=2.8$ (where $\Delta BIC_{21}=BIC_2-BIC_1$, referring to Model 2 and Model 1, respectively) deemed Model 2 not to be significantly better than Model 1.\footnote{$BIC=-2\ ln\ \mathcal{L}_{max}+k\ ln\ N$, where $\mathcal{L}_{max}$ represents the maximum likelihood that the observed data are described by the adopted model; $k$ is the number of free parameters in the model, $N$ is the number of data points. In general, $\Delta \chi^2=-2\ \Delta ln\ \mathcal{L}_{max}$, so we can calculate $BIC\approx \chi^2+k\ ln\ N$. The difference between $BIC$ values from different models, defined as $\Delta BIC=BIC_i-BIC_{min}$, can be used as evidence against the model with higher $BIC$. For instance, if $\Delta BIC>2$ there is positive evidence against model $i$, and if $\Delta BIC>6$ the evidence against model $i$ is strong (e.g., \citealt{kass1995}). The advantage of using the Bayesian information criterion is that it can be adopted to compare any kind of models, even non-nested models; however, it tends to penalize the complexity of the models.} 
Fixing $\Gamma=1.8$ provides a similarly good fit ($\chi^2/d.o.f.=94.4/101$), but with larger absorption ($N_{\rm H}=(6.5^{+0.8}_{-0.7})\times10^{23}$ cm$^{-2}$) and a high level of reflection that contrasts, however, with the weakness of the iron emission line (e.g., \citealt{walton2014}), which is not constrained in this fit. 

Using the \xmmn\ data, we jointly fit the PN and MOS spectra in the energy range $E=0.5-8$ keV (i.e. $1.5-24$ keV, rest frame, considering $z\approx2$). The spectral parameters obtained from Model 1 are: $\Gamma=1.66^{+0.27}_{-0.26}$ and $N_{\rm H}=(5.6^{+1.2}_{-1.1})\times10^{23}$ cm$^{-2}$ with redshift $z=2.04^{+0.07}_{-0.06}$ ($\chi^2/d.o.f.=144.5/147$). In the \xmm\ spectra the equivalent width of the iron emission line is EW=$136^{+106}_{-106}$ eV (rest frame), weaker than that found from the \ch\ spectrum, but consistent given the large uncertainties. The redshift in this fit is mainly constrained through the iron K edge (e.g., \citealt{iwasawa2012}). 
Fitting the spectra using Model 2, we obtain $\Gamma=1.69^{+0.44}_{-0.20}$ and $N_{\rm H}=(5.7^{+2.0}_{-1.2})\times10^{23}$ cm$^{-2}$ at $z=2.04^{+0.06}_{-0.06}$ ($\chi^2/d.o.f.=144.5/146$). If the spectral slope is fixed to $\Gamma=1.8$, we obtain an equally good fit, with consistent spectral parameter constraints. The \xmm\ data provide similar constraints as \ch\ on the spectral parameters, such as the photon index $\Gamma$, which is in agreement within the errors with the intrinsic spectral slope found for AGN (see Sect. \ref{lit}) and the column density, which proves NuSTAR J033202--2746.8 to be heavily obscured. However, as with the \ch\ spectrum, the reflection component is not constrained in the \xmmn\ spectra and according to the $BIC$, there is positive evidence against Model 2, as compared to Model 1 ($\Delta BIC_{21}=5.0$). 
The fluxes obtained from the \ch\ and \xmmn\ spectra are also consistent with each other: $f_{\rm 2-10\ keV}=(1.8^{+0.2}_{-0.1})\times10^{-14}$ \cgs, and $f_{\rm 2-10\ keV}=(1.7^{+0.2}_{-0.6})\times10^{-14}$ \cgs, respectively.

\subsection{Joint fits of \nus, \ch\ and \xmm\ spectra}\label{joint}
\begin{figure}
\centerline{
\includegraphics[scale=0.44]{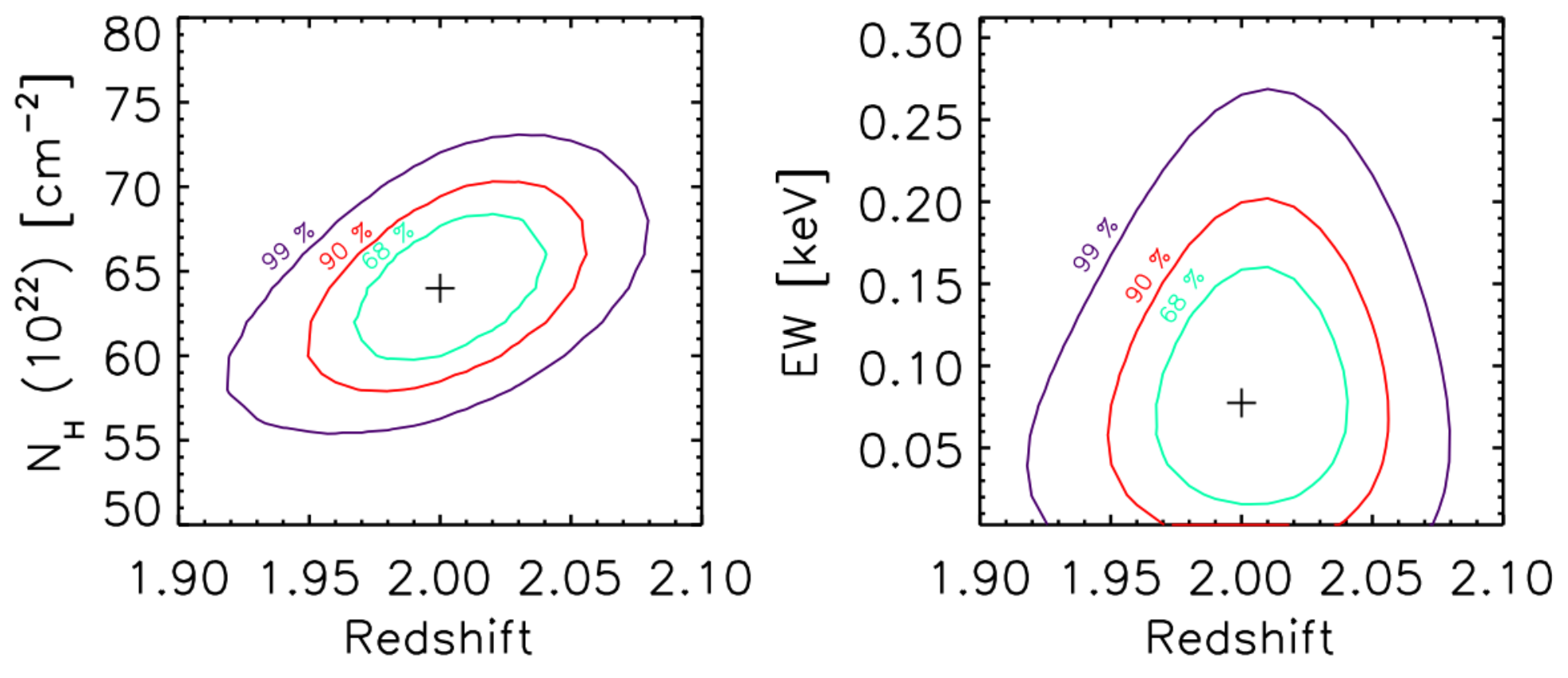}}
\caption{Redshift vs. hydrogen column density (left) and vs. Fe K$\alpha$ emission line equivalent width (rest-frame; right) contour plots obtained from the joint fit of the \nus, \xmmn, and \ch\ spectra using Model 2 with $\Gamma=1.8$; the contours correspond to 68\%, 90\% and 99\% confidence level. The redshift is well constrained in the X-ray spectra by the iron edge and the iron emission line.}
\label{fig.cont}
\end{figure}

To have a full view of the broad-band X-ray spectrum of NuSTAR J033202--2746.8, we jointly fit the \nus\ spectra with the deep lower-energy data from \ch\ and \xmmn. Although the \ch\ and \xmm\ spectra have good counting statistics, we fit the source spectra and background using C-stat, grouping the data with at least one count per bin. This is because, due to different levels of background in each dataset, we cannot use the same binning for all of the spectra (i.e., with the same number of net counts per bin). On the other hand, fitting the different spectra using different binnings for each data set can possibly cause each spectrum to have a different ``weight'' on the fit (when using $\chi^2$ statistics), biasing our best-fitting solutions.
\begin{figure}
\includegraphics[scale=0.33]{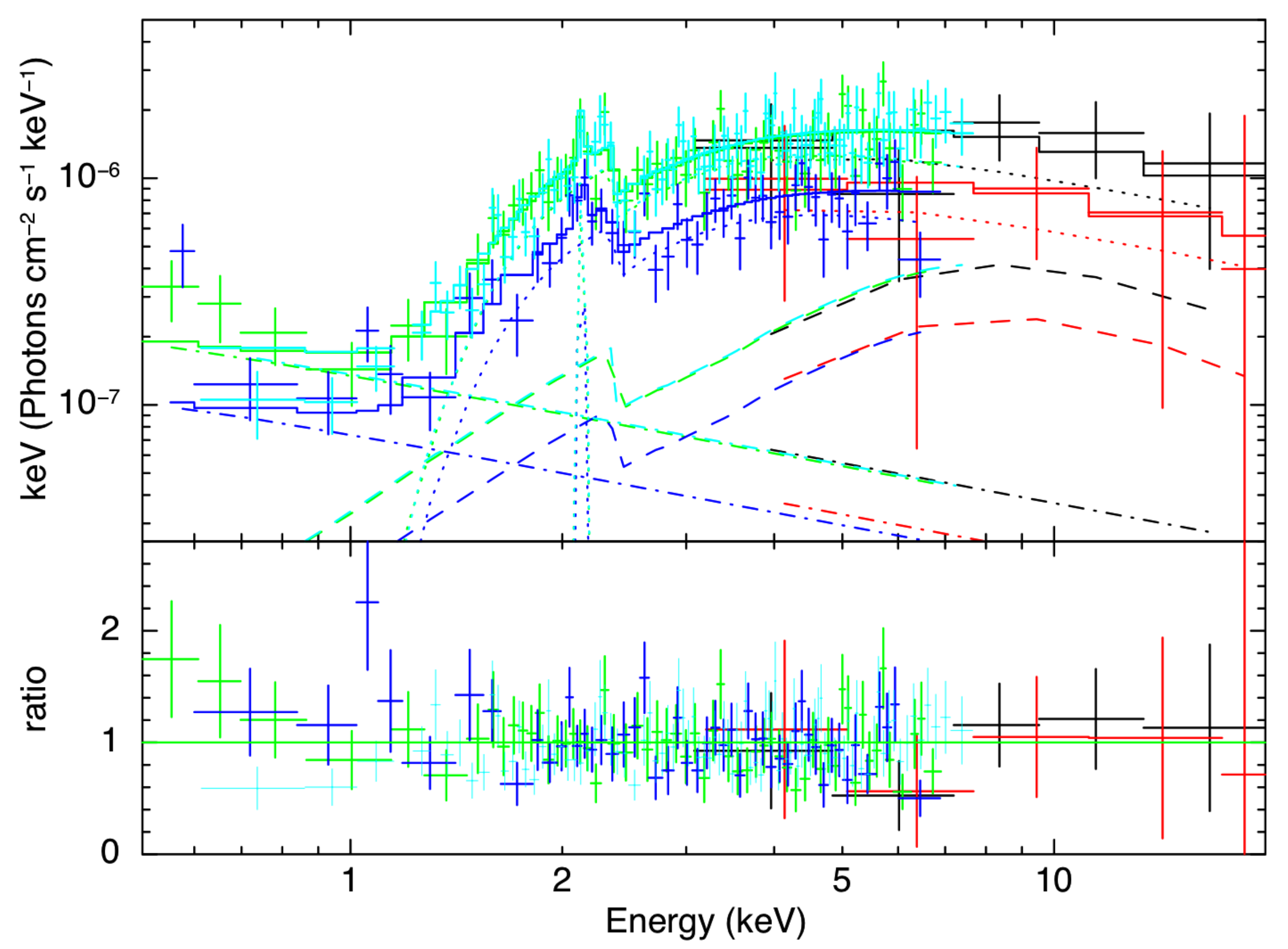}
\caption{\nus\ FPMA (black) and FPMB (red) unfolded spectra, jointly fitted with the \xmm\ PN (green) and MOS (blue) and \ch\ (cyan) spectra using Model 2 with $\Gamma=1.8$ (solid curve; see Table \ref{tab.mo}). The different components of the model are shown: transmitted absorbed power law and iron line (dotted curves), soft scattered component (dashed-dotted lines) and reflection hump (dashed curves). The spectra are background subtracted and re-binned for presentation purposes. The bottom panel shows the ratio between the data and the model.}
\label{fig.fit}
\end{figure}

We jointly fit the \nus\ spectra together with the \ch\ and \xmmn\ spectra using the models described in Sect. \ref{lowxspec} and summarized in Table \ref{tab.mo}. We introduce a third model to test whether the spectrum could be purely reflection dominated, i.e. no transmitted component is included, which would be the case if the source were heavily Compton-thick (i.e. $N_{\rm H}\gtrsim10^{25}$ cm$^{-2}$; e.g., \citealt{malizia2009, balocovic2014}). This third model is composed of a reflection component and an iron emission line, modified by absorption from material intervening our line of sight (e.g., from gas in the host galaxy), and a soft component, with spectral slope linked to the reflection component (Model 3; Table \ref{tab.mo}). Different relative normalization factors are used in the models (free to vary in the fits) to account for the cross-calibration of the different instruments. The \ch\ and \xmm\ spectra are fitted between $0.5-8$ keV (observed frame) while the \nus\ spectra are fitted between $3-30$ keV (observed frame). For the \nus\ spectra we also include a model to reproduce the background spectrum (see Sect. 4.1), fixing all of the parameters to the best-fitting values found previously (Sect. \ref{nuspec}), while we did not include a model for the \ch\ background, since it contributes very little to the total spectral counts. To fit the \xmm\ data using C-stat, however, we have to also account for the \xmm\ background, which has a significant contribution to the total counts in the spectra ($\approx50$\%). Similarly to the approach taken for the \nus\ background spectra (Sect. \ref{nuspec}), we separately fit the \xmm\ EPIC-PN and EPIC-MOS background spectra using $\chi^2$ statistics with a binning of at least 20 cts/bin to find the best-fitting parameters for the background models (e.g., \citealt{katayama2004}). We then jointly fit the \nus, \ch\ and \xmm\ spectra using the models listed in Table \ref{tab.mo}, including the relative background models with all of the parameters fixed to their best-fitting values. The resulting best-fitting parameters are reported in Table \ref{tab.sp}. Using the three datasets together we obtain tighter constraints on the redshift of the source ($z=2.00^{+0.04}_{-0.04}$) and on the spectral parameters than those obtained from individual datasets in Sect. \ref{lowxspec} (Figure \ref{fig.cont}). In Figure \ref{fig.fit} we show the results of the spectral fit using Model 2 (see Sect. \ref{sim}). The aperture-corrected flux measured from the spectra (e.g., from Model 2) in the overlapping band $E=3-8$ keV are: $f_{\rm 3-8\ keV}=(1.1\pm0.1)\times10^{-14}$ \cgs\ (\nus-FPMA), $f_{\rm 3-8\ keV}=(1.2_{-0.1}^{+0.0})\times10^{-14}$ \cgs\ (\xmm-PN) and $f_{\rm 3-8\ keV}=(1.2_{-0.1}^{+0.0})\times10^{-14}$ \cgs\ (\ch) respectively, in good agreement with each other within $\sim10$\%. In the hard \nus\ band, $E=8-24$ keV, the measured flux is $f_{\rm 8-24\ keV}=(2.7_{-0.3}^{+0.1})\times10^{-14}$ \cgs\ (\nus-FPMA). Since we verified in the previous section that the intrinsic spectral slope of NuSTAR J033202--2746.8 is consistent with $\Gamma\approx1.8$, we also fit these models fixing the photon index to this value; this increases the estimates of the column density to $N_{\rm H}\approx6.4\times10^{23}$ cm$^{-2}$ (see Table \ref{tab.sp}).

\subsection{Constraining the Compton reflection}\label{sim}
\begin{figure}
\includegraphics[scale=0.33]{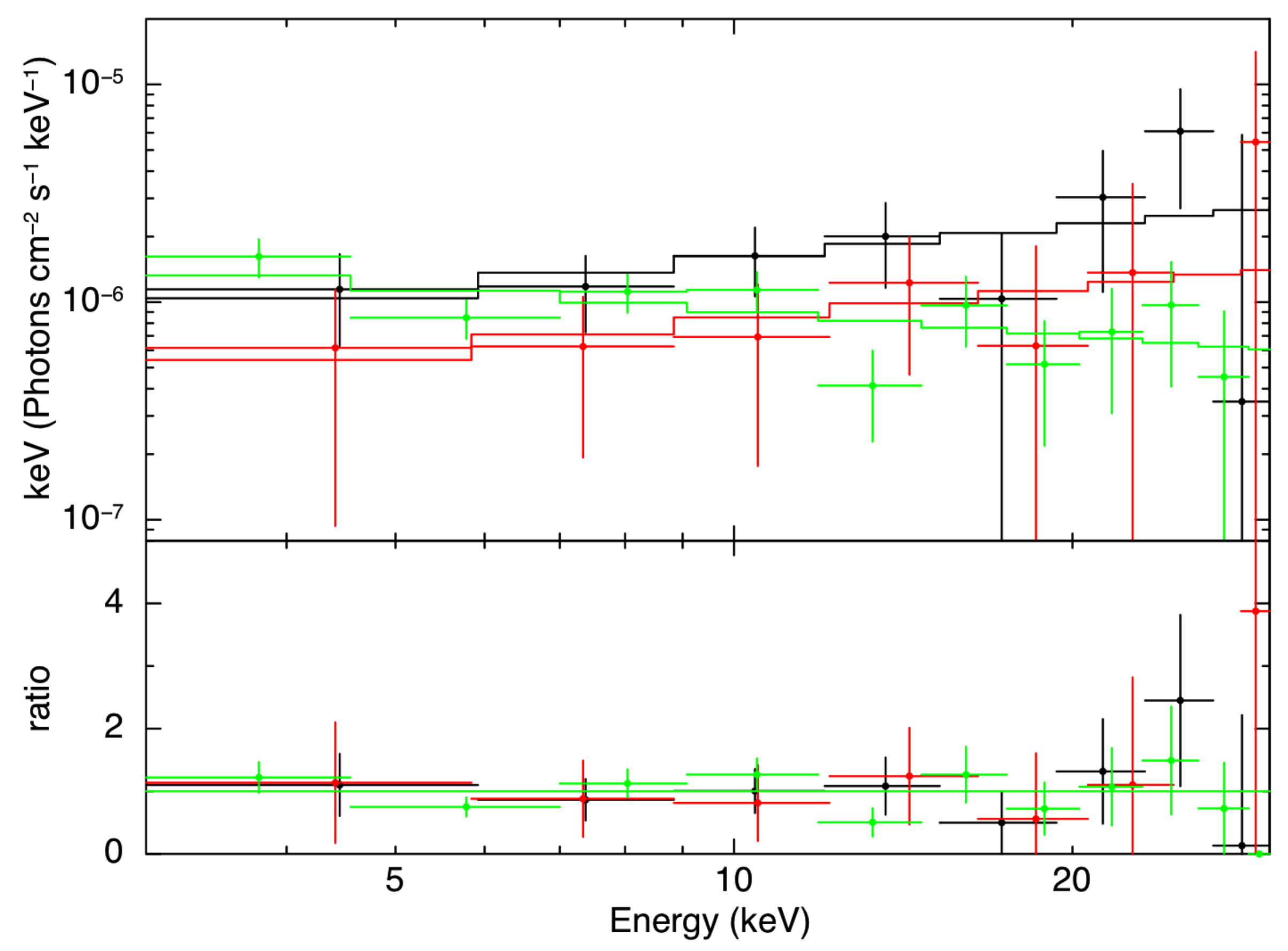}
\caption{Comparison between the \nus\ simulated spectrum (green), obtained assuming the best-fitting parameters resulting from fitting the \xmmn\ spectra with Model 1, and the real \nus\ data for NuSTAR J033202--2746.8 (FPMA: black, and FPMB: red); the spectra have been re-binned for presentation purposes. The solid lines represent the power-law models fitted to the data. The bottom panel shows the ratio between the model and the data. The real \nus\ spectra of NuSTAR J033202--2746.8 have a flatter spectral slope than that of the simulated spectrum, suggesting the need for a reflection component. }
\label{fig.sym}
\end{figure}

The limitation of using C-stat in our joint spectral fits of the \nus, \ch\ and \xmmn\ data is that with C-stat it is not possible to infer the goodness-of-fit simply through the fit-statistic values and thus we cannot significantly favor one model over the others. We therefore use spectral simulations to identify the best model to reproduce the properties of NuSTAR J033202--2746.8, and in particular, to explore whether the reflection component is required. We use the best-fitting parameters obtained from the \xmmn\ spectra using Model 1 (i.e., without a reflection component) to extrapolate the source spectrum in the \nus\ energy band and compare the characteristics of the simulated spectrum with the real \nus\ data (both FPMs; Fig. \ref{fig.sym}). We only simulate the spectrum for one of the \nus\ modules (e.g., FPMA), as the simulated spectrum for the other module is bound to be the same for a fixed model (within the uncertainties due to the instrumental response). The band ratio obtained from the simulated spectrum, $\rm CR(8-24)/CR(3-8)\approx0.86$, is lower than that measured from the real \nus\ data, even taking into account the large uncertainties on the ratio (Sect. \ref{det}). Fitting the simulated and the real spectra with a simple power-law model we obtain a flatter photon index from the real \nus\ spectra (jointly fitting FPMA and FPMB spectra) than that predicted by our simulated spectrum: $\Gamma=0.52^{+0.58}_{-0.60}$ compared to $\Gamma_{\rm sim}=1.38^{+0.29}_{-0.28}$, which are in disagreement at the 90\% confidence level. These results suggest that the reflection is probably affecting the spectrum NuSTAR J033202--2746.8 in the \nus\ band, and thus a reflection component is favored in our spectral models to best represent the observed data at $E\gtrsim10$ keV. Therefore, we identify Models 2 or 3 as best-fitting models for our source. 

From a comparison of the results obtained from the joint-fit analysis using Models 2 and 3 (see Sect. \ref{joint} and Table \ref{tab.sp}), we can rule out Model 3 as the best-fitting model. Indeed, if the source spectrum were purely reflection dominated, and hence heavily Compton-thick ($N_{\rm H}\gtrsim10^{25}$ cm$^{-2}$), the equivalent width of the iron K$\alpha$ line is expected to be much higher, EW$\gtrsim1$ keV (e.g., \citealt{ghisellini1994,levenson2002}), while the measured EW from this model is relatively low ($\sim0.2$ keV, considering the fits with fixed $\Gamma=1.8$). 

Assuming Model 2 as our best-fitting model, we can then constrain the amount of reflection in the spectrum of NuSTAR J033202--2746.8 from the ratio of the normalization of the reflected component and the transmitted power-law component. We estimate $R=0.55^{+0.44}_{-0.37}$ (for $\Gamma=1.8$). We also tested the spectral fit of the \nus, \xmmn\ and \ch\ spectra using the MYTorus model (\citealt{yaqoob2012}), which has a self-consistent treatment of the Compton scattering and fluorescent emission lines, and is a more physically motivated model than the XSPEC model components used in Model 2 (see Table \ref{tab.mo}). The results of this test-spectral fit are consistent with those obtained from our Model 2. However, we stress that the complexity of the MYTorus model is more appropriate for fitting spectra with higher counting statistics than the data presented in this paper. 
The observed X-ray luminosity of NuSTAR J033202--2746.8 derived from our preferred model (Model 2) in the rest-frame $2-10$ keV energy band is $L_{\rm 2-10\ keV}\approx1.0\times10^{44}$ \ergs\ ($L_{\rm 2-10\ keV}\approx4.0\times10^{44}$ \ergs, corrected for absorption); while the luminosity in the very hard band, at the peak of the Compton reflection hump ($10-40$ keV, rest-frame), is $L_{\rm 10-40\ keV}\approx6.4\times10^{44}$ \ergs, with $\sim$30\% of it coming from reflection ($L_{\rm 10-40\ keV,\ Refl}\approx1.8\times10^{44}$ \ergs).

\section{IR SED of NuSTAR J033202--2746.8}\label{ir}

\begin{figure}
\includegraphics[scale=0.42]{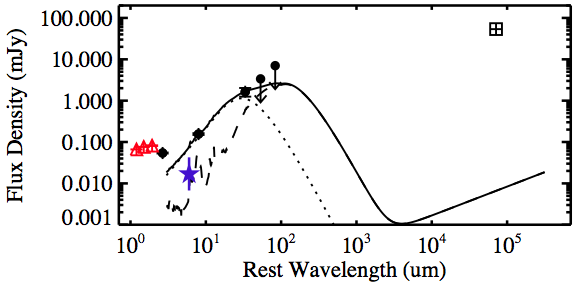}
\caption{Infrared SED of NuSTAR J033202--2746.8, using {\it Spitzer} 8 and 24 $\mu$m data and {\it Herschel} 100, 160 and 250 $\mu$m data (black points) from the GOODS-H (\citealt{elbaz2011}), PEP (\citealt{magnelli2013}) and HerMES surveys (\citealt{oliver2012}). In red triangles the {\it Spitzer} 3.6, 4.5 and 5.8 $\mu$m photometric points are also shown, while the open square represents the VLA radio flux density; these data points are not used to constrain the SED. The blue star represents the 6$\mu$m luminosity predicted from the measured X-ray luminosity ($2-10$ keV) assuming the $L_{\rm 6\mu m}-L_{\rm X}$ relation by \citet{lutz2004}. The best-fitting SED (solid curve), the AGN component (dotted curve), and the star-formation component (dashed curve) are also shown. We note that at the excess of flux density at $\lambda<3\ \mu$m (red triangles) is probably due to emission from old stellar populations in the host galaxy. The SED fit was done fixing $z=2.00$ (see Sect. \ref{joint}).}
\label{fig.sed}
\end{figure}
To obtain an independent estimate of the AGN luminosity, as well as investigate the host galaxy properties of NuSTAR J033202--2746.8, we perform a detailed SED decomposition to disentangle the contribution from the AGN and from star formation to the total mid- and far-IR emission. We use the infrared data from {\it Spitzer} at 8 and 24 $\mu$m and from {\it Herschel} at 100, 160 and 250 $\mu$m from the GOODS-H \citep{elbaz2011} and PEP catalogues (\citealt{magnelli2013}), the AGN and star-forming galaxy templates from \citet{mullaney2011}, extended to 3 $\mu$m and the radio band by \citet{delmoro2013}. The SED fitting technique is described in detail in \citet{delmoro2013}. We fixed the redshift to $z=2.00$, as measured from the X-ray spectra. The \spz\ 3.6, 4.5 and 5.8 $\mu$m data are not included in the fit as these bands ($\sim1.2-1.9\ \mu$m, rest frame at $z=2.00$) are likely to be dominated by the emission from starlight, which is not accounted for in our templates. Moreover, they fall out of the wavelength range covered by the SED templates adopted here. We find a significant AGN component ($>$99\% confidence level) dominating the IR emission up to $\approx40-50\ \mu$m (rest frame), while the cold dust emission from star formation likely dominates at longer wavelengths (see Fig. \ref{fig.sed}). From the AGN component constrained from our best-fitting SED, we measure the 6 $\mu$m luminosity of the AGN as $\nu L_{\rm 6\ \mu m}\approx3.5\times10^{44}$ \ergs, which gives an indication of the intrinsic AGN power of NuSTAR J033202--2746.8, since the extinction affecting the IR band is typically very small. {\citet{georgantopoulos2013} estimated the AGN 12 $\mu$m luminosity $\nu L_{\rm 12\ \mu m}\approx3.6\times10^{44}$ \ergs\ for this source, from their optical/IR SED decomposition. From our best-fitting SED $\nu L_{\rm 12\ \mu m}\approx8\times10^{44}$ \ergs, which is higher than that estimated by \citet{georgantopoulos2013}. We note, however, that these authors only used data up to 24 $\mu$m (8 $\mu$m, rest frame) and therefore they have no constraints on the mid- and far-IR SED of the source beyond that wavelength. The X-ray luminosity inferred from the mid-IR luminosity assuming the intrinsic $L_{\rm 6\mu m}-L_{\rm X}$ relation found for AGN (e.g., \citealt{lutz2004,fiore2009,gandi2009}) is $L_{\rm X}\approx1.4\times10^{44}$ \ergs, which is consistent, within the scatter of the correlation, with that measured from the X-ray spectra at $2-10$ keV (Sect. \ref{sim}). This supports the X-ray spectral results that our quasar, although heavily obscured, is not Compton-thick, otherwise we would expect a much lower X-ray luminosity compared to the IR one (e.g., \citealt{alexander2008b}). 

From the far-IR emission we can also place some constraints on the properties of the host galaxy of NuSTAR J033202--2746.8. Using the best-fitting SED solution we calculate the broad-band IR luminosity of the star-formation component ($8-1000\ \mu$m, rest-frame) to estimate the star-formation rate (SFR) of this source, using the \citet{kennicutt1998} relation and assuming a \citet{salpeter1955} initial mass function (IMF). We obtain SFR$\approx71\ \rm M_{\odot}~yr^{-1}$. However this has to be considered as an upper limit, since the 160 and 250 $\mu$m flux densities are upper limits and we do not have any photometric constraint beyond the SED peak (Fig. \ref{fig.sed}). \citet{georgantopoulos2013} estimated the galaxy mass (M$_*$) through broad-band optical/IR SED decomposition and reported a value of $\rm log\ (M_*/M_{\odot})=10.52$ for NuSTAR~J033202--2746.8 (assuming $z=2.0$). Although their SED analysis of this source is weakly constrained in the IR band, in the optical/near-IR bands the SED is well constrained from a large number of photometric data points, and therefore their measured stellar mass can be reliable. Using their stellar mass value for NuSTAR J033202--2746.8 we calculate the specific SFR (sSFR; i.e. the star-formation rate per unit stellar mass): sSFR$\lesssim2.1$ Gyr$^{-1}$, in agreement with the values expected for main sequence star-forming galaxies at redshift $z\approx2$ (e.g., \citealt{noeske2007,elbaz2011,mullaney2012}). However the estimated sSFR is an upper limit, implying the host galaxy of NuSTAR J033202--2746.8 might be forming stars at a smaller rate than typical star-forming galaxies. Indeed, as noted in Sect. \ref{lit}, NuSTAR J033202--2746.8 is also a bright radio source, with radio emission largely in excess of that expected from star formation (Fig. \ref{fig.sed}). Assuming $z=2.00$ we calculate the total radio luminosity $L_{\rm 1.4\ GHz}=1.2\times10^{27}$ W~Hz$^{-1}$ (rest-frame). As suggested by the radio morphology (see Sect. \ref{lit}), the radio emission is lobe-dominated (i.e., the radio core contributes $<50$\% of the total emission; e.g., \citealt{wills1995,millerb2011}), so we estimated the radio-loudness parameter $R_X=\log(\nu L_{\rm 1.4\ GHz}/L_{\rm 2-10\ keV})<-1.7$ (for $L_{core}<0.5\ L_{\rm 1.4\ GHz}$; $R_X=-2.7$ if $L_{core}=0.05\ L_{\rm 1.4\ GHz}$), which is typical of radio-loud AGN ($R_X>-2.9$; e.g., \citealt{panessa2007,tozzi2009}). Radio-excess and radio-loud AGN have been found to have smaller sSFRs than typical star-forming galaxies, or X-ray selected AGN hosts (e.g., \citealt{delmoro2013,hardcastle2013}); this effect could be related to different stages of the black hole-galaxy evolution possibly when star formation is shutting down due to AGN feedback.

\section{Discussion}\label{discus}
NuSTAR J033202--2746.8 is the highest redshift ($z\approx2$), heavily obscured quasar identified by \nus\ to date. Its hard \nus\ band ratio and the faintness in the UV, optical and near-IR bands (bluewards of the $K$-band), as well as its bright radio luminosity, make this source peculiar. 
The non-detection of the UV/optical continuum in our follow-up Keck and VLT observations is not surprising, as a relatively small amount of obscuration, coupled with the k-correction, is enough to suppress most of the emission in the UV and optical bands. The lack of emission lines in the near-IR XSHOOTER spectrum (see Sect. \ref{opt}), however, is puzzling. Taking into account the intrinsic X-ray luminosity of NuSTAR J033202--2746.8 ($L_{\rm 2-10\ keV}\approx4.0\times10^{44}$ \ergs) and the $L_{\rm [OIII]}-L_{\rm X}$ and $L_{\rm H\alpha}-L_{\rm X}$ relations from \citet{panessa2006}, we would expect $f_{\rm[OIII]}\approx3\times10^{-16}$ \cgs and $f_{\rm H\alpha}\approx4\times10^{-16}$ \cgs, while from the spectrum we measure line flux upper limits that are $>$30 times smaller. This discrepancy is still significant if we account for the large scatter of the $L_{\rm [OIII]}-L_{\rm X}$ and $L_{\rm H\alpha}-L_{\rm X}$ relations. Possible explanations for the lack of the emission lines could be: i) the source redshift: at $z\approx1.95-2.0$ (i.e., within the errors of our redshift estimates from the X-ray spectra) the [OIII]$\ \lambda 5007$ \AA\ and H$\alpha\ \lambda6563$ \AA\ emission lines are shifted to the observed wavelengths where there is no, or little atmospheric transmission; ii) obscuration on large scales: in obscured (i.e., type 2) AGN the obscuration is typically attributed to material surrounding the nuclear black hole. This material occults the UV/optical/soft X-ray continuum emission from the black hole and the broad emission lines, which are emitted from gas in the vicinity of the nucleus (e.g., \citealt{antonucci1993}), but does not affect the narrow emission lines, which are emitted on larger scales. However, if a significant amount of obscuring material is present on large scales, e.g., in the narrow line regions (NLRs) or in the host galaxy (e.g., \citealt{brand2007}), the emission from the narrow lines can also be reddened or suppressed. Both scenarios could be consistent with the broad-band properties of NuSTAR J033202--2746.8, however, it is not possible to favor one of them with the current data. 

From the broad-band X-ray coverage of \nus, together with \ch\ and \xmmn\ data, we were able to fully characterize the X-ray spectrum of NuSTAR J033202--2746.8. This source is obscured by high column density $N_{\rm H}\approx6\times10^{23}$ cm$^{-2}$ material, a factor of $\sim2-3$ higher than what was found in previous works (e.g., \citealt{castellomor2013,georgantopoulos2013}). The intrinsic power-law slope $\Gamma\approx1.6$ is consistent with the typical spectra of unobscured AGN within errors ($\Gamma\approx1.8$), but, in particular, there is good agreement with spectral slopes typically seen in radio-loud AGN, which are somewhat flatter than in radio-quiet AGN (e.g., \citealt{pagek2005}; although, see also \citealt{sambruna1999}). X-ray spectral characteristics very similar to NuSTAR J033202--2746.8 have been observed in the local Universe for 4C$+$29.30, also identified as a heavily obscured radio-loud quasar \citep{sobolewska2012}. The Fe K$\alpha$ emission line identified in the \ch\ and \xmm\ spectra, which allowed us to determine the redshift of the source together with the iron K edge, is slightly weak (EW$\approx140$ eV) when compared to the expectations for heavily obscured (Compton-thin) AGN (EW$\approx200-300$ eV, for the inclination angle assumed in our models; e.g., \citealt{ghisellini1994}). However, the strength of the iron line depends on many parameters, such as the source inclination angle, the torus opening angle and the spectral index of the underlying continuum (e.g., \citealt{george1991,ghisellini1994,levenson2002,nandra2007}). Moreover, a weakening of the iron emission line has been observed in radio-loud quasars (e.g., \citealt{reeves2000}). \citet{reeves2000} suggest that the weakening of the Fe K$\alpha$ emission line, as well as the flattening of the spectral slope, correlate with the radio-loudness parameter and depend on the increasing of the Doppler boosting of the X-ray continuum when the radio jet angle approaches the line of sight. This effect also suppresses the Compton reflection hump. 
In the soft band ($E<2$ keV) the spectrum of NuSTAR J033202--2746.8 is dominated by a scattered component that we parameterize with a power law, whose fraction is $f_{scatt}\sim4$\% of the primary (transmitted) power law (see Table \ref{tab.sp}); this emission is often seen in Seyfert 2 galaxies and quasars and is generally attributed to: 1) scattering of the primary emission by hot gas (e.g., \citealt{matt1996}), 2) ``leakage'' of a fraction of the nuclear emission due to partial covering of the central black hole (e.g., \citealt{vignali1998, corral2011}), or 3) emission from a circumnuclear starburst, or star formation in the galaxy (e.g., \citealt{maiolino1998b}). The soft component is typically only a few percent of the primary power law, though in some cases where the nucleus is heavily buried in a geometrically thick cold gas torus (with a solid angle $>2\pi$) the scattered fraction can be very small ($f_{scatt}<0.5$\%; e.g., \citealt{ueda2007,comastri2010}). In radio-loud quasars the soft X-ray emission is found to correlate with the radio-core luminosity (e.g., \citealt{worrall1994,hardcastle1999,evans2006,hardcastle2006}; \citealt{millerb2011}), suggesting that the soft X-rays in these sources are related to the relativistic jets, and might originate at the base of the radio jets. Due to the complex radio morphology observed for NuSTAR J033202--2746.8 (Sect. \ref{lit}; \citealt{miller2013}), however, we do not have a measurement of the radio-core luminosity of our source and cannot verify the correlation between the soft X-rays and radio emission. Therefore we are not able to unambiguously assess the origin of the $E<2$ keV emission in NuSTAR J033202--2746.8. 

In the hard band ($E>10$ keV) the spectrum of NuSTAR J033202--2746.8 shows indications of a Compton reflection component with a relative normalization of $R=0.55^{+0.44}_{-0.37}$, which was not constrained in previous studies of \ch\ and \xmmn\ data alone. Although this component is relatively weak (and our constraints have large scatter), it is consistent with the measured strength of the iron K$\alpha$ line, according to the relation found in well studied local sources (e.g., \citealt{walton2014}). The amount of reflection in NuSTAR J033202--2746.8 is possibly larger than that observed in bright quasars ($R\lesssim0.3$). In particular, it is somewhat larger than that found for radio-loud quasars, where the reflection is typically $R<<1$, or consistent with no reflection (e.g., $R<0.1$; \citealt{reeves2000}). This suggests that in NuSTAR J033202--2746.8 the Doppler-boosted jet component is likely not dominating over the reflection component (see also, \citealt{sobolewska2012}), and therefore its spectrum is in some aspects more similar to those of radio-quiet quasars. Our analyses of the broad-band X-ray spectrum of NuSTAR J033202--2746.8 testifies to the importance of \nus\ high energy data to fully characterize the spectral properties of sources out to high redshift. Although radio-loud AGN constitute only a small fraction of the total AGN population, NuSTAR J033202--2746.8 demonstrates that there can be a range of properties in the X-ray spectra of AGN, such as the amount of reflection in bright quasars, that needs to be quantified, and perhaps accounted for in our population synthesis models. This is very important because having better constraints on the spectra of individual sources is essential to improve our models of the XRB. These models are largely affected by degeneracies in their many parameters (e.g., \citealt{gandi2007,treister2009}), such as the source spectral models, luminosity functions, and column density distribution of AGN, amongst others. Different assumptions on these ``ingredients'' can have a significant impact on the predictions we extract from these models and on our understanding of the AGN population and its space density (e.g., \citealt{comastri1995,ueda2003,treister2005,gilli2007,treister2009,ballantyne2011}).  
On these matters great progress will be made in the near future, since large AGN samples are now available with \nus\ in the E-CDF-S and COSMOS fields (Mullaney et al., and Civano et al., in prep.), and accurate broad-band X-ray spectral analysis of these sources will allow us to place constraints on the shape of the AGN spectra up to high energies. 

\section{Conclusions}\label{concl}
We performed detailed X-ray spectral analysis of NuSTAR J033202--2746.8, the source with the highest band ratio found in the \nus\ observations of the E-CDF-S field so far. The source is very faint in the optical band ($R>25.5$ mag), indicating significant reddening, and bright in the radio band ($L_{\rm 1.4\ GHz}\approx1.2\times10^{27}$ W~Hz$^{-1}$). Using the \nus\ hard X-ray data  in combination with existing deep \ch\ and \xmmn\ data, we investigate the broad-band X-ray properties of NuSTAR J033202--2746.8. Moreover, using deep mid- and far-IR data we perform SED decomposition to fully characterize the multi-wavelength properties of the source. 
Our results can be summarized as follows: 
\begin{itemize}
\item[--] follow-up UV-to-near-IR spectroscopy reveals a faint continuum in the near-IR band with no detection at shorter wavelengths, supporting the idea that the source is obscured in the optical bands. No emission line is detected in the spectrum, preventing a redshift identification from these spectra (however flux losses, and atmospheric absorption might have affected our results). We are planning to perform further follow-up observations, e.g. with the {\it Hubble Space Telescope} (HST) to avoid the atmospheric transmission issues. \smallskip
\item[--] Although no secure redshift identification is available from optical/near-IR spectroscopy for the source, we constrain the redshift from the X-ray spectra: $z=2.00\pm0.04$, in agreement with \citet{georgantopoulos2013}. The X-ray luminosity estimated from the X-ray spectra is $L_{\rm 2-10\ keV}\approx10^{44}$ \ergs\ ($L_{\rm 2-10\ keV}\approx4\times10^{44}$ \ergs, corrected for absorption), and $L_{\rm 10-40\ keV}=6.4\times10^{44}$ \ergs, around the peak of the Compton reflection. \smallskip
\item[--] From the broad-band X-ray spectral analysis we constrain NuSTAR J033202--2746.8 to be heavily obscured with a column density $N_{\rm H}\approx6\times10^{23}$ cm$^{-2}$, $\sim2-3$ times higher than that previously found using \ch\ or \xmm\ data alone (e.g., \citealt{tozzi2006,castellomor2013,georgantopoulos2013}). \smallskip
\item[--] By jointly fitting the \nus, \xmm\ and \ch\ data, and using spectral simulations, we find indications of a Compton reflection component contributing $\sim30$\% to the total emission of NuSTAR J033202--2746.8 at $E\approx10-40$ keV (rest frame), and we estimate the reflection fraction $R=0.55^{+0.44}_{-0.37}$; although this component is relatively weak, it is stronger than that previously found for bright radio-loud quasars, whose X-ray spectra are typically consistent with no reflection.\smallskip 
\item[--] The IR SED analysis reveals the mid-IR emission of NuSTAR J033202--2746.8 is dominated by an AGN component, with $\nu L_{\rm 6\ \mu m}\approx3.5\times10^{44}$ \ergs, in agreement with the AGN power estimated in the X-rays, while the far-IR SED is possibly dominated by cool dust emission due to star formation; however, we only place an upper limit on the specific SFR$\lesssim2.1$ Gyr$^{-1}$, which could be consistent with typical star-forming galaxies at $z\approx2$, but also with the lower sSFRs observed in radio-excess and radio-loud AGN.
\end{itemize}
Although NuSTAR J033202--2746.8 shows some peculiar characteristics, such as the extreme X/O flux ratio (see Sect. \ref{lit}), the lack of optical/near-IR emission lines (Sect. \ref{opt}), and the hard \nus\ band ratio, we conclude that they can be explained through large amount of obscuration around the central black hole or on larger scales. The X-ray spectral properties of NuSTAR J033202--2746.8 are not peculiar, or rare, and could be fairly typical in quasars. We have shown that having higher energy data ($E>10$ keV) is essential to provide a full characterization of the spectral properties of the source, especially at $E\approx20-30$ keV, at the peak of the Compton reflection hump. Such a full spectral characterization is often not feasible when only using lower energy data ($E<10$ keV), but it is essential to make progress on our understanding of the XRB composition and on the AGN population.
\acknowledgements
We thank the anonymous referee for careful reading and for the helpful comments, which helped improving this manuscript. We gratefully acknowledge financial support from the UK Science and Technology Facilities Council (STFC, ST/I001573/I, ADM and DMA; ST/K501979/1, GBL; ST/J003697/1, PG) and the Leverhulme Trust (DMA and JRM). AC, CV, RG, and PR thank the ASI/INAF grant I/037/12/0 - 011/13. FEB acknowledges support from Basal-CATA (PFB-06/2007) and CONICYT-Chile (FONDECYT 1101024 and Anillo grant ACT1101) and ET acknowledges the FONDECYT grant 1120061. WNB and BL thank Caltech \nus\ subcontract 44A-1092750 and NASA ADP grant NNX10AC99G. MB acknowledges the International Fulbright Science and Technology Award. This work was supported under NASA Contract No. NNG08FD60C, and made use of data from the \nus\ mission, a project led by the California Institute of Technology, managed by the Jet Propulsion Laboratory, and funded by the National Aeronautics and Space Administration. We thank the \nus\ Operations, Software and Calibration teams for support with the execution and analysis of these observations. This research has made use of the \nus\ Data Analysis Software (NuSTARDAS) jointly developed by the ASI Science Data Center (ASDC, Italy) and the California Institute of Technology (USA). This work also used observations made with ESO Telescopes at the La Silla Paranal Observatory under the programme ID 092.A-0452.

\bibliographystyle{apj}

\end{document}